\documentclass{article}

\usepackage[english]{babel}

\usepackage[letterpaper,top=2cm,bottom=2cm,left=3cm,right=3cm,marginparwidth=1.75cm]{geometry}

\usepackage{amsmath}
\usepackage{graphicx}
\usepackage{algorithm}
\usepackage{algorithmic}
\usepackage{bm}
\usepackage{tabularx}
\usepackage{multirow, multicol}
\usepackage{longtable}
\usepackage{comment}
\usepackage{subfigure}
\usepackage[colorlinks=true, allcolors=blue]{hyperref}

\addto\extrasenglish{}
\addto\extrasenglish{}

\def\0{\bf 0}

\def\E{\bf E}

\def\O{\bf O}

\newcolumntype{C}{>{\centering\arraybackslash}X}

\title{Improving Disease Risk Estimation in Small Areas by Accounting for spatiotemporal Local Discontinuities}
\author{Guzmán Santafé, Aritz Adin, M. Dolores Ugarte}
\date{}

\begin{document}
\maketitle

\begin{abstract}

This work proposes a two-step method to enhance disease risk estimation in small areas by integrating spatiotemporal cluster detection within a Bayesian hierarchical spatiotemporal model. First, we introduce an efficient scan-statistic-based clustering algorithm that employs a greedy search within the scan window, enabling flexible cluster detection across large spatial domains. We then integrate these detected clusters into a Bayesian spatiotemporal model to estimate relative risks, explicitly accounting for identified risk discontinuities. We apply this methodology to large-scale cancer mortality data at the municipality level across continental Spain.
Our results show our method offers superior cluster detection accuracy compared to SaTScan. Furthermore, integrating cluster information into a Bayesian spatiotemporal model significantly improves model fit and risk estimate performance, as evidenced by better DIC, WAIC, and logarithmic scores than SaTScan-based or standard BYM2 models. This methodology provides a powerful tool for epidemiological analysis, offering a more precise identification of high- and low-risk areas and enhancing the accuracy of risk estimation models.

\end{abstract}

\section{Introduction}

In recent years, small-area disease risk estimation has gained increasing attention due to its crucial role in epidemiological research and public health decision-making. In disease mapping, two primary approaches are typically followed: one focuses on cluster detection to identify regions with significantly high or low risk (or rates), while the other aims at accurately estimating disease risks. 
For cluster detection, window-based scan statistics methods are commonly used, with popular options including SaTScan and FlexScan. SaTScan \cite{kulldorf1997} allows for identifying both high and low-risk clusters in spatial and spatio-temporal settings. In the spatial case, its scanning window is either circular or elliptical, while in the spatio-temporal case, it employs a cylindrical scanning window \cite{kulldorf1998}. This constraint makes SaTScan less suitable when the true risk surface contains arbitrarily shaped clusters. In contrast, FlexScan \cite{tango2005} allows for the detection of arbitrarily shaped clusters but only for purely spatial problems. To address cluster detection in a spatio-temporal context, \cite{takahashi2008} extends the flexible scan window by proposing a prospective search for pyramidal clusters with an arbitrarily shaped spatial base. While this method is more flexible than SaTScan for spatio-temporal cluster detection, it is also limited to very specific contexts and clustering shapes. Unfortunately, its implementation is not included in the FlexScan software, and, to the best of our knowledge, it is not publicly available.

For the objective of relative risk estimation, statistical models incorporating main spatial and temporal random effects as well as spatio-temporal interactions   \cite{besag1991,RueHeld2005,riebler2016} are widely used. However, the smoothing induced by these models can obscure discontinuities and mask the clusters present on the risk surface, potentially leading to misleading interpretations.
Several studies have sought to bridge the gap between cluster detection and risk estimation by integrating cluster information into spatio-temporal models, thereby capturing local discontinuities and achieving a better balance between smoothing and cluster preservation. For instance, recent studies have proposed the use of clustering algorithms and Bayesian hierarchical models together to better estimate local risks.
Anderson et al. \cite{anderson2014} develop a Bayesian clustering approach to improve disease mapping by identifying spatial clusters in the first stage and incorporating this information into the modeling framework in the second stage of their proposed method. Adin et al. \cite{adin2019} extend this two-stage methodology to spatio-temporal models with conditional autoregressive (CAR) priors, enabling better risk estimation in the presence of local discontinuities. Similarly, Santafé et al. \cite{santafe2021} address this issue in a purely spatial context by proposing a new density-based clustering algorithm that facilitates the analysis of large areas, demonstrating improvements in cancer mortality risk estimation. More recently, Yin et al. \cite{yin2022} explore the use of multiple clustering algorithms to generate different clustering partitions, which were then used to derive new neighborhood matrices incorporated into a Bayesian spatio-temporal disease mapping model as a prior uniform discrete distribution over the neighborhood matrix. Similarly, \cite{Yin2025} employ a graph-based optimization algorithm to estimate a set of candidate neighborhood matrices. 
All of these proposals, however, fail to consider the data distribution when detecting clusters. 
%

In this work, we adopt the integrated approach described above by proposing a methodology that first identifies significant spatio-temporal clusters of both high and low risk, and then incorporates this information into a Bayesian hierarchical spatio-temporal risk estimation model. To account for the underlying probability distribution of the observed data, we propose \textit{GscanStat}, a novel and efficient scan-statistic-based method that uses a greedy search within the scan window, facilitating flexible cluster detection in large spatial domains. Finally, we incorporate the detected clusters into a Bayesian spatio-temporal model to accurately estimate risks while accounting for the discontinuities identified by the clustering partition. To evaluate the effectiveness of the proposed methodology, we conduct an extensive simulation study comparing \textit{GscanStat} against SaTScan with a cylindrical scan window. We use SaTScan as a reference since it is the publicly available state-of-the-art method for spatio-temporal clustering detection. The comparison is performed in terms of cluster detection accuracy, risk estimation bias and variance, and model selection criteria. Additionally, we apply the proposed methodology to a large-scale real-world case study to analyze spatio-temporal patterns and identify significant risk clusters using cancer mortality data at the municipal level across continental Spain.

The remainder of this paper is organized as follows. \autoref{sec:Method} describes the \textit{GscanStat} algorithm and the hierarchical Bayesian model used for disease risk estimation. \autoref{sec:simulation.study} presents the simulation study, including data generation, clustering performance evaluation, and model assessment. \autoref{sec:case.study.SP} provides a case study on cancer mortality risk in a large-scale case considering cancer mortality of all municipalities in continental Spain, highlighting the practical application of our approach. Finally, \autoref{sec:conclusions} summarizes the main conclusions and discusses potential future research directions.

\section{Cluster detection and disease risk estimation}
\label{sec:Method}

This section introduces the proposed two-stage approach for cluster detection and risk smoothing in spatio-temporal disease mapping. The main contribution of this work comes in the first stage of the proposed method. Specifically, we introduce \textit{GscanStat}, an efficient clustering algorithm based on scan statistics.  Its core functionality is the detection of spatio-temporal aggregations characterized by statistically significant high or low relative risk.

Let $A_i^t$ represent the  $i$-th spatial unit  ($i=1,\ldots, n$) within the study region at time period $t$  ($t=1,\ldots, T$). Let
${\O}=(O_1^1,\ldots, O_n^1,\ldots, O_1^T,\ldots, O_n^T)'$ and ${\E}=(E_1^1,\ldots, E_n^1,\ldots, E_1^T,\ldots, E_n^T)'$ denote the observed and expected number of disease cases for each area and time period, respectively. The expected cases are typically calculated using internal age and sex standardization. The spatio-temporal clustering algorithm detects arbitrarily shaped groups of neighboring areas. An area's neighbors are either physically adjacent, sharing a common border (spatial neighbors), or they represent the same location observed across consecutive time intervals (temporal neighbors).
  The goal is to identify clusters where the total number of observed disease cases is significantly higher or lower than the expected number of cases. The proposed \textit{GscanStat} algorithm is empirically evaluated in a simulated scenario and compared to previously proposed spatio-temporal clustering methods for disease mapping, such as the SaTScan method.

\subsection{GscanStat: A spatio-temporal clustering algorithm based on scan statistics}
\label{sec:clustering-alg}

The \textit{GscanStat} algorithm is a greedy search method that automatically detects spatio-temporal clusters of any shape. The proposed method resembles other scan statistic methods, such as the cylindrical space-time scan statistic by \cite{kulldorf1998} (SaTScan) and the prismatic space-time scan statistic by \cite{takahashi2008}. Similar to the space-time scan statistic framework under the Poisson model, we assume that, conditional on the relative risk $r_i^t$, the number of observed cases in area $i$ at time $t$ ($O_i^t$) follows a Poisson distribution. Specifically,
\begin{equation}
\label{poisson.process}
    O_i^t \mid r_i^t \sim \text{Poisson}(\mu_i^t = r_i^t \cdot E_i^t),
\end{equation}

\noindent where $r_i^t$ denotes the relative risk of mortality (or incidence) for area $i$ at time $t$ ($A_i^t$). As is common in this setting, we assume that under the null hypothesis of no space-time clusters, all relative risks are equal to 1, i.e.,
\begin{equation*}
H_0: r_i^t = 1, \quad \forall \,\, i, t.
\end{equation*}

\noindent A limiting cylindrical window ($W^*$) is imposed around each area $A_i^t$. This window includes areas $A_{i'}^{t'}$ such that $A_{i'}$ is among the $K$ nearest neighbors of $A_i$ in space and $t' \in \left[\max(1,t-T^*), \min(T,t+T^*) \right]$, where $K$ and $T^*$ denote the maximum spatial and temporal window sizes, respectively, both set as algorithm parameters.
Let $W$ be a flexible-shaped window on the area $A_i^t$, defined as a subset of $W^*$. This subset is constructed such that all areas in $W$ are spatially or temporally connected to $A_i^t$. Additionally, $\mathcal{W}$ denotes the class of all possible flexible-shaped windows within $W^*$. Thus, considering a common relative risk within and outside a window $W$, we have 
%
%
\begin{eqnarray*}
    r_i^t =
    \begin{cases}
        r(W) & \text{if } A_i^t \in W \\
        r(W^c) & \text{otherwise},
    \end{cases}
\end{eqnarray*}

\noindent where $W^c$ denotes all areas except those in $W$. By reformulating the null hypothesis within the framework of a space-time scan statistic test for cluster detection, we can write 
\begin{eqnarray*}
    H_0: r(W) = r(W^c). 
\end{eqnarray*}
The alternative hypothesis depends on whether we aim to detect high or low-risk clusters. Therefore $H_1: r(W) > r(W^c)$ is used for hot-spot detection and $H_1: r(W) < r(W^c)$ for cold-spot detection. The test statistics for hot-spot ($\lambda_H$) and cold-spot ($\lambda_C$) detection are given by
\begin{eqnarray*}
    \lambda_H = \sup_{W\in \mathcal{W}} \left(\frac{O(W)}{E(W)}\right)^{O(W)} \left(\frac{O - O(W)}{O-E(W)}\right)^{O-O(W)} I\left( \frac{O(W)}{E(W)} > \frac{O-O(W)}{O-E(W)} \right), \\[2.ex]
     \lambda_C = \sup_{W\in \mathcal{W}} \left(\frac{O(W)}{E(W)}\right)^{O(W)} \left(\frac{O - O(W)}{O-E(W)}\right)^{O-O(W)} I\left( \frac{O(W)}{E(W)} < \frac{O-O(W)}{O-E(W)} \right),
\end{eqnarray*}

\noindent where $I(\cdot)$ is the indicator function, $O(W)$ and $E(W)$ denote the observed and expected number of cases within the window $W$ under the null hypothesis, and $O = \sum_{i=1}^n\sum_{t=1}^T O_i^t$ is the total number of cases in the study region. The windows $W_H$ and $W_C$, for which $\lambda_H$ and $\lambda_C$ are obtained, identify the most likely high-risk and low-risk clusters, respectively. The test statistic distributions under the null hypothesis for high/low risk clustering detection are obtained by Monte Carlo hypothesis testing. Note that, in our case, two independent one-tailed tests are conducted to detect clusters rather than a two-tailed test, because our simulations empirically show that the probability distributions of $\lambda_H$ and $\lambda_C$ differ. Therefore, using a two-tailed test might hinder the identification of low-risk clusters.
Given that we permit completely arbitrary cluster shapes, the number of potential scanning windows could be exceedingly large, rendering the evaluation of every single window $W\in \mathcal{W}$ infeasible, even for relatively small-scale problems. Hence, we propose a greedy search strategy to approximate the most likely spatio-temporal clusters. This strategy iteratively expands the window, centered on each initial space-time location, by incorporating the spatially or temporally neighboring area that yields the greatest increase in the likelihood ratio.
The pseudo-code for the proposed greedy search algorithm is presented in Algorithm~\ref{alg:greedy-search}.

\begin{algorithm}[ht]
\renewcommand{\thealgorithm}{\arabic{algorithm}}
\begin{algorithmic}[1]
\FOR{$t=1:T$}
    \FOR{$i=1:n$}
        \IF {$O_i^t < E_i^t$}
            \STATE \# low-risk cluster
            \STATE $(W_C)_i^t = \{A_{it}\}$
        \ELSE 
           \STATE \# high-risk cluster
           \STATE $(W_H)_i^t = \{A_{it}\}$
        \ENDIF
        \WHILE{$\lambda$ increases}
        \STATE candidates = $(W_C)_i^t$ [or $(W_H)_i^t$] space/time neighbors within $\mathcal{W}$ 
        \STATE $(W_C)_i^t \text{ [ or } (W_H)_i^t \text{ ] }\leftarrow $ append candidate maximizing $\lambda$
        \ENDWHILE
    \ENDFOR
\ENDFOR
\end{algorithmic}
\caption{\label{alg:greedy-search}GscanStat greedy search algorithm.}
\end{algorithm}

The most likely low-risk cluster (MLCC) is the $(W_C)_i^t$ with the highest $\lambda_C$  value. Similarly, the most likely high-risk cluster (MLCH) is the $(W_H)_i^t$ with the highest $\lambda_H$ value. Based on the probability distribution of the test statistics $\lambda_C$ and $\lambda_H$ obtained through Monte Carlo simulations, we determine whether MLCC and MLCH are statistically significant. Additionally, significant secondary clusters beyond the most likely cluster can be identified and ranked according to their likelihood ratio test statistic ($\lambda$ value).

\subsection{Disease risk estimation model}
\label{sec:models}
Given a clustering configuration $\mathbf{C} = (C_1, \ldots, C_c)$ obtained using the GscanStat algorithm described in \autoref{sec:clustering-alg}, which incorporates the most likely significant high- and low-risk clusters together with significant secondary clusters, a Bayesian hierarchical model is fitted to the data. Conditional on the relative risk, the observed cases are assumed to follow a Poisson distribution, as stated in \autoref{poisson.process}. Therefore, the mean of the Poisson distribution is modeled as
\begin{eqnarray}
\label{eq.log-risk}
    \log \mu_i^t = \log E_i^t + \log r_i^t,
\end{eqnarray}

\noindent where $\log E_i^t$ is an offset, and the specification of $\log r_i^t$ depends on the clustering partition $\mathbf{C}$. If no cluster structure is detected, the log-risk in \autoref{eq.log-risk} is expressed as
\begin{eqnarray}
    \log r_i^t = \alpha + \xi_i + \gamma^t + \delta_i^t,
    \label{mod:noclusters}
\end{eqnarray}

\noindent where $\alpha$ is a global intercept, $\xi_i$ is a spatially structured random effect, $\gamma^t$ is a time-structured random effect, and $\delta_i^t$ is a spatio-temporal random effect. 
For the spatial random effect $\boldsymbol{\xi} = (\xi_1, \ldots, \xi_n)'$, we assume a BYM2 prior distribution \cite{riebler2016}. Namely
\begin{equation}
\boldsymbol{\xi} \sim \mathcal{N}(0, \mathbf{Q}_{\xi}^*) \quad \text{with} \quad \mathbf{Q}_{\xi}^* = \tau^{-1}_{\xi}\left[(1 - \lambda_{\xi}) \mathbf{I}_n + \lambda_{\xi} \mathbf{R}_*^{-}\right]
\end{equation}

\noindent where $\tau_{\xi}$ denotes the precision parameter, $\lambda_{\xi} \in [0, 1]$ denotes a spatial smoothing parameter, $\mathbf{I}_n$ is the $n \times n$ identity matrix, and $\mathbf{R}_*^{-}$ is the Moore–Penrose generalized inverse of the scaled spatial precision matrix associated with the undirected graph representing the study region \cite{Sorbye2014}.

The temporally structured random effect is denoted as $\bm{\gamma} = (\gamma^1, \dots, \gamma^T)$. This effect is modeled assuming a first-order random walk (RW1) prior distribution:  
\begin{equation*}
\bm{\gamma} \sim N \left( \bm{0}, [\tau_{\gamma} \bm{R}_{\gamma}]^{-} \right),
\end{equation*}

\noindent where \( \tau_{\gamma} \) represents a precision parameter, \( \bm{R}_{\gamma} \) is the \( T \times T \) structure matrix associated with the RW1 process \cite{RueHeld2005}, and $^-$ stands for the Moore-Penrose generalized inverse.  

Finally, for the space-time interaction effect $\bm{\delta} = (\delta_{1}^1, \dots, \delta_{n}^1, \dots, \delta_{1}^T, \dots, \delta_{n}^T)$, we assume the following distribution 
\begin{equation*}
\bm{\delta} \sim N \left( \bm{0}, [\tau_{\delta} \bm{R}_{\delta}]^{-} \right),
\end{equation*}

\noindent where \( \tau_{\delta} \) is a precision parameter, and \( \bm{R}_{\delta} \) is an \( nT \times nT \) matrix derived as the Kronecker product of the spatial and temporal structure matrices. Four alternative specifications for \( \bm{R}_{\delta} \), corresponding to different types of space-time interaction structures, can be considered \cite{KnorrHeld2000} (see \autoref{tab:interactions_constraints}). 

We note that the proposed model (\autoref{mod:noclusters}) suffers from identifiability issues. Therefore, we impose sum-to-zero constraints on the random effects to ensure our estimates are well-defined.  \autoref{tab:interactions_constraints} summarizes the appropriate constraints for each interaction type, following \cite{goicoa2018}.

\begin{table}[ht]
\renewcommand{\arraystretch}{1.5}
\centering
\resizebox{\textwidth}{!}{
\begin{tabular}{lccc}
\hline
Interaction & \( \bm{R}_{\delta} \) & Correlation  & Constraints \\
\hline
Type I   & \( \bm{I}_T \otimes \bm{I}_n \) & None & \( \sum_{i=1}^{n} \xi_i = 0, \quad \sum_{t=1}^{T} \gamma^t = 0, \quad \sum_{i=1}^{n} \sum_{t=1}^{T} \delta_{i}^t = 0 \) \\
Type II  & \( \bm{R}_{\gamma} \otimes \bm{I}_n \) & Temporal & \( \sum_{i=1}^{n} \xi_i = 0, \quad \sum_{t=1}^{T} \gamma^t = 0, \quad \sum_{t=1}^{T} \delta_{i}^t = 0, \forall i \) \\
Type III & \( \bm{I}_T \otimes \bm{R}_{\xi} \) & Spatial & \( \sum_{i=1}^{n} \xi_i = 0, \quad \sum_{t=1}^{T} \gamma^t = 0, \quad \sum_{i=1}^{n} \delta_{i}^t = 0, \forall t \) \\
Type IV  & \( \bm{R}_{\gamma} \otimes \bm{R}_{\xi} \) & Spatial and Temporal & \( \sum_{i=1}^{n} \xi_i = 0, \quad \sum_{t=1}^{T} \gamma^t = 0, \quad \sum_{i=1}^{n} \delta_{i}^t = 0, \forall t, \quad \sum_{t=1}^{T} \delta_{i}^t = 0, \forall i \) \\
\hline
\end{tabular}}
\caption{\label{tab:interactions_constraints} Different space-time interaction structures and their corresponding identifiability constraints.}
\end{table}

On the other hand, if some cluster structure is detected, i.e. $C$ contains at least one cluster, the log-risk in \autoref{mod:noclusters} is modeled as
\begin{equation}
\log r_{i}^t = \alpha + \xi_i + \gamma^t + \delta_{i}^t + \sum_{j=1}^{c} I[A_{i}^t\in C_j] \cdot \beta_j,
\label{mod:clusters}
\end{equation}

\noindent where $\alpha, \bm{\xi}, \bm{\gamma}$, and $\bm{\delta}$ are defined as in the no-cluster model (Model~\ref{mod:noclusters}) and $\bm{\beta} = (\beta_1,\ldots, \beta_c)'$ are fixed effects associated with each cluster, with $I[\cdot]$ an indicator function so that $I[A_{i}^t\in C_j]$ equals 1 if the area $A_i^t$ lies in cluster $C_j$ and zero otherwise. 

In all the models described above, improper uniform prior distributions on the positive real line are considered for the 
standard deviations, i.e. $\sigma_e = 1 / \sqrt{\tau_e} \sim U(0,\infty)$, with $e=\xi, \gamma, \delta$. A standard uniform distribution is also given to the spatial smoothing parameters $\lambda_{\xi}$. Finally, $N(0, 10^6)$ priors are given to the fixed effects $\beta_j$.
We fitted all models using R-INLA (version 25.06.07 on R-4.5.1) together with the \texttt{bigDM} package (version 0.5.7) \cite{bigDM}, a release specifically developed for fitting the models presented in this paper.

\section{Simulation Study}
\label{sec:simulation.study}

We conduct a simulation study to compare the performance of three different models: 

\begin{enumerate}

\item The non-clustered model, as described in  \autoref{mod:noclusters}.

\item The SaTScan model, which uses cylindrical spatiotemporal clusters detected by the widely known SaTScan software, as defined in \autoref{mod:clusters}.

\item The GscanStat model, also based on \autoref{mod:clusters}, but incorporates the clusters obtained from the greedy search algorithm (see \autoref{sec:clustering-alg}).

\end{enumerate}

\noindent

\begin{figure}[ht]
  \centering
  \includegraphics[width=\textwidth]{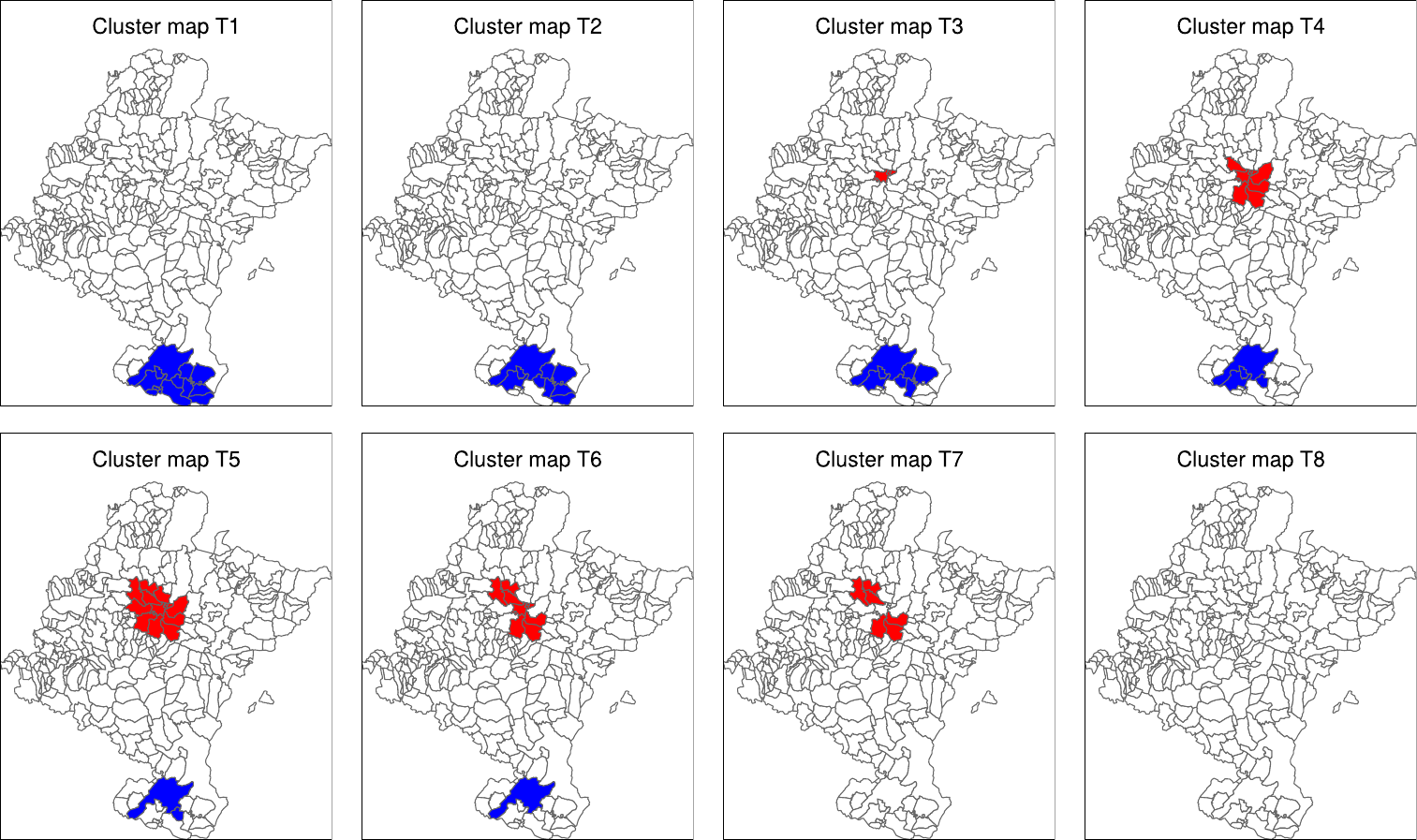}
  \caption{Spatio-temporal clusters used in the simulation study.}
  \label{fig:clusters}
\end{figure}

The region of interest for this study is the Spanish Autonomous Community of Navarre, located in northern Spain. It comprises 265 municipalities (n=265) observed over a span of eight time periods (T=8).
The simulation setting considers the two arbitrary-shaped spatiotemporal clusters presented in \autoref{fig:clusters}. Using these specific cluster partitions, we establish three different scenarios and sub-scenarios to comprehensively evaluate the performance of our proposed methodology.


\subsection{Data generation}
\label{sec:simExp_dataGeneration}

The number of observed cases ($O_i^t$ with $i=1,\ldots,256$ and $t=1,\ldots,8$) is simulated from a Poisson distribution with mean $E_i^t\cdot r_i^t$, where $E_i^t$ denotes the expected number of cases. 
We compute the expected cases from global cancer mortality data for the Navarre region, aggregated into eight 3-year periods spanning 1999–2022. They range from 0.1 to 1003.0, with a mean of 9.8 and a median of 2.3. The risk of the $i$-th area at time $t$ ($r_i^t$) depends on the specific scenario/subscenario considered. Namely, 
\begin{itemize}
\item Scenario A includes only cluster fixed effects.

\item Scenario B incorporates spatial, temporal, and spatio-temporal random effects, with no clustering effects.  

\item Scenario C extends scenario B by including the cluster fixed effects. 

\end{itemize}
\autoref{tab:scenarios} summarizes the models for each scenario, where $\beta_j$ denotes the fixed effect for the $j$-th cluster, and $\xi_i, \gamma^t$, and $\delta_i^t$ are spatial, temporal, and spatio-temporal random effects, respectively. When clustering fixed effects are considered in the scenario, $\beta_j$ takes the value $\log(2.5)$ if $C_j$ is a high-risk cluster and $\beta_j=\log(0.4)$ if $C_j$ is a low-risk cluster. The spatial random effect vector $\bm{\xi} = (\xi_1,\ldots,\xi_n)'$ is generated from an intrinsic CAR prior distribution \cite{besag1991}. The temporally structured random effect vector $\bm{\gamma}=(\gamma^1,\ldots,\gamma^T)'$ is generated from an RW1 prior and the spatio-temporal random effect is sampled from a completely structured precision matrix (see Type IV interaction in \autoref{tab:interactions_constraints}). 
The precision parameter $\tau$ for each random effect ($\bm{\xi}, \bm{\gamma}, \bm{\delta}$) is adequately chosen so that the spatial ($\bm{\xi}$) and temporal ($\bm{\gamma}$) effects take values within the range [0.85, 1.15], while the spatio-temporal interaction ($\bm{\delta}$) lies within [0.95, 1.05]. This implies that the risk associated with an area ($\xi_i$) or a time period ($\gamma^j$) may be up to 15\% lower or higher than the overall risk, whereas the risk for a specific area at a specific time ($\delta_i^t$) varies by at most 5\%.

\begin{table}[ht]
  \centering
  \begin{tabular}{|c|c|}
    \hline
    Scenario & Model \\ \hline 
     & \\[-3mm]
    A & $\log r_i^j = \sum_{j=1}^c I(A_i^t \in C_j) \beta_j$ \\[2mm]
    B & $\log r_{i}^t = \alpha + \xi_i + \gamma^t + \delta_{i}^t$ \\ [2mm]
    C & $\log r_{i}^t = \alpha + \xi_i + \gamma^t + \delta_{i}^t + \sum_{j=1}^{c} I[A_{i}^t\in C_j] \cdot \beta_j$\\[2mm] \hline
  \end{tabular}
  \caption{Models used to simulate $r_i^t$ in each scenario.}
  \label{tab:scenarios}
\end{table}

For scenarios A and C, three different subscenarios are considered. According to \autoref{fig:clusters}, two spatio-temporal clusters are defined: one in red and one in blue. In this context, subscenario 1H considers only the red cluster as a high-risk cluster. In subscenario 1L, only a low-risk cluster is considered within the red cluster. Finally, in subscenario 1H1L, the red cluster is a high-risk cluster, while the blue cluster is a low-risk cluster. 

Scenario A is designed to evaluate the clustering detection performance of the proposed method. In contrast, scenario B tests the method's behavior when no true clusters are present. Finally, scenario C assesses the proposed algorithm in a complex setting involving spatial, temporal, and spatio-temporal correlations. 
The combination of scenarios and subscenarios yields seven distinct configurations, each used to generate 100 simulated datasets.

\subsection{Clustering detection}
\label{sec:simExp_clusterigDetection}

In this experiment, we assess the ability of the proposed methodology to accurately identify high and low-risk clusters. Specifically, we compare the results obtained using  GscanStat with those from the SaTScan algorithm employing a cylindrical search window. For the GscanStat algorithm, we consider a maximum spatial window size of $K = 150$ and a maximum temporal window size of $T = 5$. For the SaTScan clustering algorithm, we perform a retrospective space-time analysis using a circular spatial window with a maximum size of 50\% of the population at risk and a temporal window covering up to 90\% of the study period. Note that the spatial and temporal window size specifications differ between GscanStat and SaTScan. Larger window sizes cover broader areas in the clustering analysis but incur higher computational costs. Since our goal is to capture as large a region as possible, the parameter values used for SaTScan correspond to the maximum sizes allowed by the software. Nevertheless, the configuration proposed for GscanStat spans broader and more flexible spatio-temporal regions than the maximum configuration allowed by SaTScan.

In the experiments, {\it recall} quantifies the proportion of areas that truly belong to a high-risk or low-risk cluster (see \autoref{fig:clusters}) and are correctly identified as such by the clustering algorithms. A high recall means that there are few false negatives, i.e., there are few areas that belong to the cluster that were missed by the algorithm. Additionally, the {\it precision} represents the proportion of areas identified as high-risk or low-risk by the clustering algorithm that actually belong to true high-risk or low-risk clusters, respectively. A high precision means that most of the areas are assigned to the cluster they truly belong to, i.e, there are few false positives.  
Furthermore, we compute the average number of clusters identified by the clustering algorithms across the 100 simulated datasets, along with the average number of spurious clusters (i.e., clusters in which none of the areas are accurately classified as high-risk or low-risk). 

\begin{table}[h!]
\centering
\begin{tabularx}{\textwidth}{|l|>{\centering\arraybackslash}X >{\centering\arraybackslash}X >{\centering\arraybackslash}X >{\centering\arraybackslash}X |>{\centering\arraybackslash}X >{\centering\arraybackslash}X >{\centering\arraybackslash}X >{\centering\arraybackslash}X|}
\hline
\multirow{2}{*}{\textbf{Scenario}} & \multicolumn{4}{c|}{\textbf{SaTScan}} & \multicolumn{4}{c|}{\textbf{GscanStat}} \\ \cline{2-9}
        & \textbf{Recall} & \textbf{Prec.} & \textbf{\#detec. clust.} & \textbf{\#spur. clust.} 
        & \textbf{Recall} & \textbf{Prec.} & \textbf{\#detec. clust.} & \textbf{\#spur. clust.}\\ \hline
A\_1H1L     & 0.12          & 0.36          & 2.95          & 1.91
            & \textbf{0.91} & \textbf{0.65} & \textbf{2.02} & \textbf{0.02} \\ \hline
A\_1H       & 0.14          & 0.41          & 2.58          & 1.56
            & \textbf{0.96} & \textbf{0.53} & \textbf{1.05} & \textbf{0.05} \\ \hline
A\_1L       & 0.00          & 0.00          & 3.04          & 3.03
            & \textbf{0.92} & \textbf{0.49} & \textbf{1.03} & \textbf{0.03} \\ \hline
B           &  --           &  --           & 2.91          & 2.91
            &  --           &  --           & \textbf{1.52} & \textbf{1.52} \\ \hline
C\_1H1L     & 0.13          & 0.37          & 2.92          & 1.92
            & \textbf{0.75} & \textbf{0.50} & \textbf{2.59} & \textbf{0.59} \\ \hline
C\_1H       & 0.14          & 0.42          & 2.64          & 1.63
            & \textbf{0.65} & \textbf{0.34} & \textbf{1.80} & \textbf{0.80} \\ \hline
C\_1L       & 0.01          & 0.00          & 4.01          & 3.68
            & \textbf{0.65} & \textbf{0.31} & \textbf{1.69} & \textbf{0.69} \\ \hline

\end{tabularx}
\caption{Clustering detection: Recall, Precision (Prec.), the average number of detected clusters (\#detec. clust.) and average number of spurious clusters (\#spur. clust.) in each one of the scenarios/subscenarios.}
\label{tab:clustering_detection}
\end{table}

As shown in \autoref{tab:clustering_detection}, the GscanStat method consistently outperforms SaTScan in terms of recall across all scenarios, indicating that GscanStat more effectively identifies the true spatio-temporal clusters in the actual risk surface. In scenario A, the simplest case for clustering detection, GscanStat achieves recall values exceeding 0.9 and correctly identifies both high-risk and low-risk clusters. In contrast, SaTScan yields very low recall values and fails to detect low-risk clusters (with a recall of 0.00 in scenario $A\_1L$). In scenario C, which presents complex risk surfaces sampled from a model with spatial, temporal, and spatio-temporal random effects, as well as fixed clustering effects, GscanStat also achieves high recall values (greater than 0.65 in all subscenarios), clearly outperforming SaTScan. In terms of precision, GscanStat also achieves higher values than SaTScan. However, these values are not particularly high, indicating that GscanStat also tends to detect larger clusters than the actual ones.   

When analyzing the number of detected clusters, we observe that in scenario A, GscanStat provides a close estimate of the true number of clusters. As a result, the average number of spurious clusters remains close to zero. In contrast, the SaTScan algorithm identifies a substantial number of spurious clusters, particularly in scenarios involving low-risk clusters. For instance, in scenario $A\_1L$, an average of 3.04 clusters is detected, of which 3.03 are spurious. Thus, SaTScan mostly detects false high-risk clusters in this scenario. Similarly, in scenario $A\_1H1L$, SaTScan detects an average of 2.95 clusters, with 1.91 being spurious. This suggests that, on average, only 1.04 clusters correspond to true clusters, with the detected clusters corresponding exclusively to high-risk clusters.  Scenario C is much more complex than scenario A, and therefore, the average number of spurious clusters increases. Nevertheless, the number of clusters detected by GscanStat is still close to the true number of clusters, with less than 1 spurious cluster on average. By contrast, SaTScan obtains similar or even worse results than in scenario A. In scenario B, even when no clusters are present in the risk surface, both clustering methods still detect clusters. This may occur because, when sampling models with complex spatio-temporal interactions, some of these relationships may randomly appear as spatio-temporal clusters, which clustering algorithms might then overestimate. Nevertheless, GscanStat detects roughly half the average number of overestimated clusters compared to SaTScan.

\subsection{Model fitting}
\label{sec:simExp_modelFitting}

In this experiment, we compare the fitting performance of three models: the GscanStat model, the SaTScan model, and the model without clusters (noCluster). \autoref{tab:fitting} summarizes the results based on several model selection criteria: the Deviance Information Criterion (DIC) \cite{spiegelhalter2002bayesian}, the Watanabe–Akaike Information Criterion (WAIC) \cite{watanabe2010asymptotic}, and the Logarithmic Score (LS) \cite{gneiting2007strictly}, with the best values highlighted in bold.
The DIC metric consists of two components: the effective number of parameters ($p_D$), which is reported in the table, and the posterior mean of the deviance (not shown). The $p_D$ provides insight into the model's complexity and the degree of smoothing of the random effects.

Across all scenarios, GscanStat consistently outperformed the other models according to all selection criteria. In scenario B, where no space-time clusters are present, all models yield  relatively low $p_D$ values, suggesting that the spatial and temporal random effects are subject to substantial smoothing, thereby contributing minimally to the model’s overall complexity.
However, in scenarios with clusters (scenarios A and C), a clear space-time pattern associated with these clusters emerges: the GscanStat model exhibits low $p_D$ values, in contrast to the substantially higher values observed for the SaTScan and noCluster models. This difference can be attributed to GscanStat's ability to accurately identify and incorporate the underlying clustering structure of the data, as detailed in \autoref{sec:simExp_clusterigDetection}. 
By capturing the main structured variability through fixed effects associated with the identified clusters, GscanStat reduces the demand on spatial and temporal random effects, which remain smooth and contribute minimally to model complexity, resulting in lower $p_D$ values.
In contrast, the noCluster model fails to account for the presence of clusters, while the SaTScan model does not detect them accurately. As a result, much of the variability attributable to the clusters is absorbed by the spatial and spatio-temporal random effects, increasing both model complexity and the number of effective parameters. Despite this increased complexity, model selection criteria show poorer performance for these models.\\


\begin{table}[h!]
\centering
\begin{tabularx}{\textwidth}{|l|>{\centering\arraybackslash}X >{\centering\arraybackslash}X >{\centering\arraybackslash}X >{\centering\arraybackslash}X |
                                >{\centering\arraybackslash}X >{\centering\arraybackslash}X >{\centering\arraybackslash}X >{\centering\arraybackslash}X |
                                >{\centering\arraybackslash}X >{\centering\arraybackslash}X >{\centering\arraybackslash}X >{\centering\arraybackslash}X |}
\hline
\multirow{2}{*}{\textbf{Scenario}} & \multicolumn{4}{c|}{\textbf{noCluster}} & \multicolumn{4}{c|}{\textbf{SaTScan}} & \multicolumn{4}{c|}{\textbf{GscanStat}} \\ \cline{2-13}
        & $p_D$  & \textbf{DIC}  & \textbf{WAIC} & \textbf{LS}  
        & $p_D$  & \textbf{DIC}  & \textbf{WAIC} & \textbf{LS}  
        & $p_D$  & \textbf{DIC}  & \textbf{WAIC} & \textbf{LS}  \\ \hline
A\_1H1L  & 393  & 8451  & 8464  & 4335  
         & 335  & 8431  & 8459  & 4288  
         & \textbf{30}  & \textbf{8030}  & \textbf{8029}  & \textbf{4015}  \\ \hline
A\_1H    & 351  & 8426  & 8431  & 4294  
         & 281  & 8399  & 8422  & 4249  
         & \textbf{32}  & \textbf{8078}  & \textbf{8077}  & \textbf{4039}  \\ \hline
A\_1L    & 281  & 8278  & 8282  & 4180  
         & 257  & 8270  & 8278  & 4169  
         & \textbf{25}  & \textbf{7968}  & \textbf{7967}  & \textbf{3983}  \\ \hline
B        & 36   & 8116  & 8115  & 4058  
         & \textbf{35}   & 8108  & 8107  & 4054  
         & 48   & \textbf{8024}  & \textbf{8021}  & \textbf{4013}  \\ \hline
C\_1H1L  & 390  & 8435  & 8446  & 4321  
         & 332  & 8412  & 8438  & 4275  
         & \textbf{58}  & \textbf{7999}  & \textbf{7997}  & \textbf{4001}  \\ \hline
C\_1H    & 351  & 8431  & 8437  & 4295  
         & 279  & 8403  & 8426  & 4250  
         & \textbf{10}  & \textbf{8034}  & \textbf{9552}  & \textbf{4063}  \\ \hline
C\_1L    & 278  & 8267  & 8271  & 4173  
         & 241  & 8253  & 8263  & 4158  
         & \textbf{42}  & \textbf{7917}  & \textbf{7915}  & \textbf{3958}  \\ \hline
\end{tabularx}
\caption{Simulated data: average value of the effective number of parameters ($p_D$), deviance information criterion (DIC), Watanabe-Akaike information criterion (WAIC), and logarithmic score (LS).}
\label{tab:fitting}
\end{table}

\clearpage
\subsection{Risk estimation}
\label{sec:simExp_riskEstimation}

To evaluate the models' performance in estimating relative risks, we compute the mean absolute bias (MAB) and the mean root mean square error (MRMSE) on the log-risk scale for each areal unit as: 
\begin{equation*}
    MAB_i = \frac{1}{T}\sum_{t=1}^T \frac{1}{100}\left|\sum_{l=1}^{100} \hat{lr}_{i}^{t}(l) - lr_i^t\right| , 
    \hspace{0.3cm} 
    MRMSE_i = \frac{1}{T}\sum_{t=1}^T \sqrt{\frac{1}{100}\left(\sum_{l=1}^{100} \hat{lr}_{i}^{t}(l) - lr_i^t\right)^2}
\end{equation*}

\noindent where \(lr_i^t\) and \(\hat{lr}_i^t(l)\) denote the true generated log-risk and the posterior median estimate, respectively, for the \(i\)-th area at time \(t\) in the \(l\)-th simulation. Coverage probabilities and the lengths of 95\% credible intervals are also computed. Additionally, we report the interval score (IS) for the 95\% credible interval of the log-risks \cite{gneiting2007strictly}, a metric that combines both interval length and empirical coverage. This enhances the interpretability of the results and facilitates model comparison. The IS for the 95\% credible interval is defined as follows
\begin{eqnarray}
    IS_{0.05}(lr) = (u-l) + \frac{2}{0.05}\left[(l-lr)I\left[lr<l\right] + (lr-u)I\left[lr>u\right]\right],
\end{eqnarray}

\noindent where \([l, u]\) is the 95\% credible interval for the log-risk, and \(I[\cdot]\) is an indicator function that penalizes the length of the credible interval if the true log-risk $lr$ is not included in the credible interval.

The overall interval score for area $i$ is also computed by averaging the median interval scores across 100 simulations and all time periods as
\begin{eqnarray}
    IS_{0.05}(lr_i) = \frac{1}{T}\sum_{t=1}^T median\left(\left\{IS_{0.05}(lr_i^t(l))\right\}_{l=1}^{100}\right)
\end{eqnarray}

\noindent where $IS_{0.05}(lr_i^t(l))$ is the interval score at simulation $l$ for area $i$ and time period $t$.
The \textit{median} is used instead of the mean to reduce the influence of outliers, which can arise due to spatio-temporal variability and randomness in the data. This approach ensures a more robust assessment of model performance.

\autoref{tab:riskEvaluation} presents the results from 100 simulations, summarizing average values across small areas for the entire risk surface. In scenarios A and C, which contain high-risk and/or low-risk clusters, the GscanStat model achieves lower bias (MAB) and variance (MRMSE) in risk estimation compared to the other models. With respect to the credible intervals for the risk estimates, GscanStat generally produces wider intervals while maintaining high coverage levels, leading to lower IS values.
In contrast, in scenario B, which assumes a homogeneous risk surface (i.e., no clusters), GscanStat performs comparably to the noClusterModel in terms of bias and variance, but slightly underperforms relative to SaTScan. 
Still, these differences are minor when contrasted with the notable improvements GscanStat demonstrates in other scenarios, outperforming both SaTScan and the noClusterModel.

\begin{table}[!h]
    \centering
    \begin{tabular}{|c|c|c|c|c|c|c|}
        \hline
        \textbf{Scenario} & \textbf{Model} & \textbf{MAB} & \textbf{MRMSE} & \textbf{Length} & \textbf{Cover95} & $\bm{IS_{0.05}}$ \\
        \hline
        \multirow{3}{*}{A-1H} 
        &  noClusterModel & 0.0430 & 0.4330 & 0.7860 & 0.9750 & 0.8550 \\ 
        &  SaTScan & 0.0360 & 0.3620 & 0.6870 & 0.9730 & 0.7830 \\ 
        &  GscanStat & \textbf{0.0180} & \textbf{0.1830} & 0.1860 & 0.9780 & \textbf{0.1810} \\         
        \hline
        \multirow{3}{*}{A-1H1L} 
        &  noClusterModel & 0.0510 & 0.5040 & 0.8570 & 0.9740 & 0.9230 \\ 
        &  SaTScan & 0.0440 & 0.4370 & 0.7770 & 0.9690 & 0.8700 \\ 
        &  GscanStat & \textbf{0.0230} & \textbf{0.2320} & 0.1790 & 0.9640 & \textbf{0.2050} \\ 
        \hline
        \multirow{3}{*}{A-1L} 
        &  noClusterModel & 0.0480 & 0.4790 & 0.7040 & 0.9710 & 0.8450 \\ 
        &  SaTScan & 0.0510 & 0.5100 & 0.6700 & 0.9680 & 0.8180 \\ 
        &  GscanStat & \textbf{0.0220} & \textbf{0.2190} & 0.1730 & 0.9760 & \textbf{0.2000} \\ 
        \hline
        \multirow{3}{*}{B} 
        &  noClusterModel & 0.0290 & 0.2690 & 0.2060 & 0.9860 & 0.2080 \\ 
        &  SaTScan & \textbf{0.0250} & \textbf{0.2360} & 0.2080 & 0.9550 & \textbf{0.2070} \\ 
        &  GscanStat & 0.0290 & 0.2670 & 0.2360 & 0.9280 & 0.2420 \\         \hline
        \multirow{3}{*}{C-1H} 
        &  noClusterModel & 0.0490 & 0.4880 & 0.7860 & 0.9750 & 0.8550 \\ 
        &  SaTScan & 0.0450 & 0.4450 & 0.6860 & 0.9710 & 0.7830 \\ 
        &  GscanStat & \textbf{0.0350} & \textbf{0.3510} & 0.2470 & 0.9270 & \textbf{0.2490} \\ 
        \hline
        \multirow{3}{*}{C-1H1L} 
        &  noClusterModel & 0.0570 & 0.5660 & 0.8550 & 0.9730 & 0.9210 \\ 
        &  SaTScan & 0.0530 & 0.5280 & 0.7740 & 0.9680 & 0.8680 \\ 
        &  GscanStat & \textbf{0.0410} & \textbf{0.4090} & 0.2620 & 0.9350 & \textbf{0.2840} \\ 
        \hline
        \multirow{3}{*}{C-1L} 
        &  noClusterModel & 0.0540 & 0.5430 & 0.7020 & 0.9690 & 0.8480 \\ 
        &  SaTScan & 0.0560 & 0.5590 & 0.6450 & 0.9610 & 0.8170 \\ 
        &  GscanStat & \textbf{0.0400} & \textbf{0.4010} & 0.2270 & 0.9220 & \textbf{0.2620} \\ 
        \hline
    \end{tabular}
    \caption{\label{tab:riskEvaluation} Log-risk estimation performance for the whole region under study across 100 simulations.}
\end{table}

Alternatively, we also evaluate the models' risk estimation ability on specific regions of the risk surface, such as the areas included in true high-risk or low-risk clusters. This is of great interest for understanding how these models behave.
\autoref{tab:riskEvaluation_HLClust} summarizes the performance metrics for areas belonging to high-risk clusters (top) and low-risk clusters (bottom) across all scenarios and subscenarios. In the first setting, the GscanStat model achieves the best values across all scores, demonstrating the proposed method’s ability to more accurately identify high-risk clusters and better estimate the risk within these regions.
In low-risk subscenarios (1L, 1H1L), GscanStat again outperforms the other models in low-risk cluster areas. However, it is worth noting that the GscanStat model performs worse in low-risk areas than in high-risk areas, indicating that low-risk scenarios pose greater challenges than high-risk ones.
Finally, \autoref{tab:riskEvaluation_noCluster} presents results for areas outside the clusters, reflecting the overall risk surface evaluation. GscanStat again performs better than both SaTScan and the noCluster models, although in this case the differences are less pronounced than those observed in areas belonging to high-risk and low-risk clusters.

These results demonstrate that the proposed method effectively identifies high-risk and low-risk clusters with high precision, and that this information is appropriately integrated into the risk estimation model to yield more accurate estimates.

\section{Case study: Overall cancer mortality in Spain}
\label{sec:case.study.SP}


This section analyzes the spatio-temporal risk estimates of overall cancer (all sites) mortality among males in continental Spain from 1999 to 2022. The study area comprises 7,906 municipalities.
Although the data are aggregated into 3-year periods (T = 8 time points), many small municipalities exhibit very low observed and expected death counts, introducing substantial variability that can hinder the performance of the GscanStat algorithm in detecting clusters.
To mitigate this issue, we apply a preprocessing step that merges neighboring municipalities, while respecting the administrative boundaries of the Autonomous Communities, until each new spatial unit (aggregated municipalities) contains at least 16 observed cases over the entire study period. This approach not only reduces the substantial variability of the SMRs but also enhances the reliability of the final risk estimates \cite{quick2024reliable}. It is important to note that, assuming a Poisson distribution for the observed counts, a minimum threshold of sixteen cases per area ensures a coefficient of variation below 0.25 for the SMR.

After aggregation, the final configuration comprises 2,470 regions across continental Spain. Despite the reduction in the number of small areas, fitting spatio-temporal models with complex interaction effects remains computationally intensive, even with the efficiency gains afforded by approximate Bayesian methods like INLA.
Therefore, in this study, we adopt the recently proposed divide-and-conquer strategy by \cite{orozco2023} implemented in the R package \texttt{bigDM} \cite{bigDM}, which reduces the computational challenges posed by high-dimensional areal data while enabling the fitting of local spatio-temporal models that more accurately capture underlying data patterns.

\begin{figure}[!ht]
    \centering
    \vspace{0.1cm}
    \includegraphics[width=10cm]{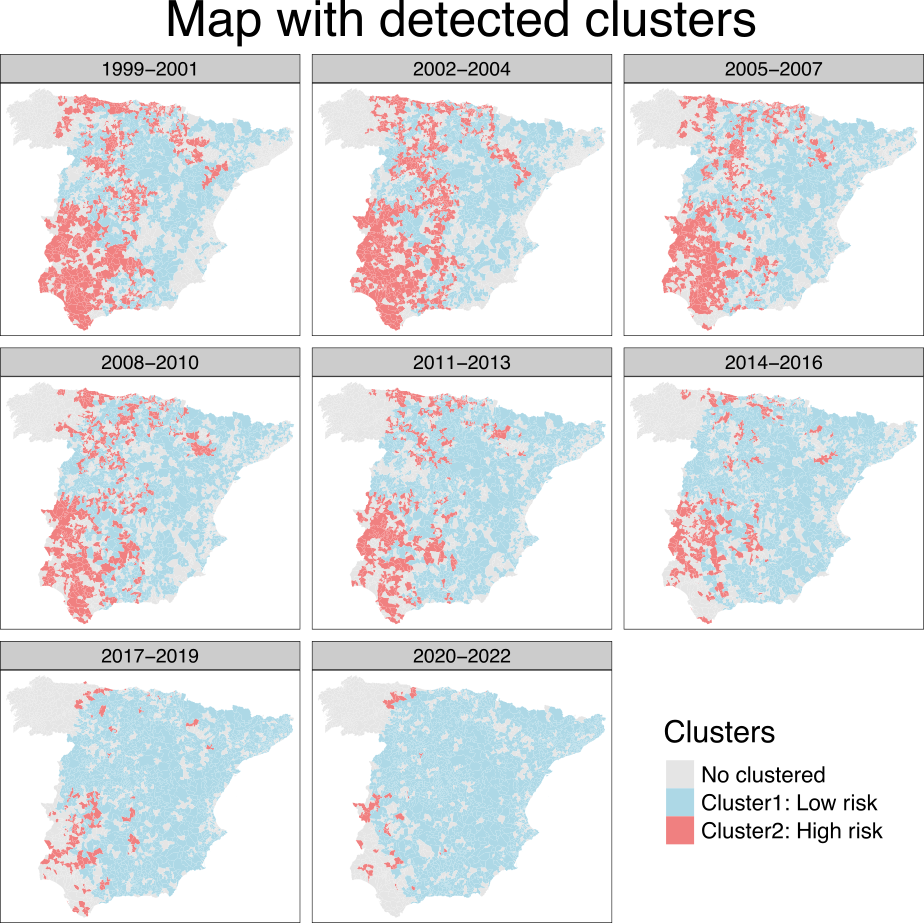}    
    \caption{Clusters identified by the GscanStat algorithm in the analysis of overall cancer mortality in Spain.}
    \label{fig:SP_clusterPartition}
\end{figure}

First, the GscanStat clustering algorithm is applied to the entire study region (comprising 2,470 areas over 8 time periods) using a maximum spatial window size of $K=2000$ and a maximum temporal window size of $T=5$. The algorithm requires approximately 12 hours to generate the clustering partition on an Intel(R) Xeon(R) Silver 4316 processor with 80 CPUs at 2.30 GHz and 256 GB of RAM. However, as the algorithm is fully parallelized, this runtime can be significantly reduced by allocating additional computational resources if necessary. The resulting partition reveals two significant clusters: a high-risk cluster and a low-risk cluster (see \autoref{fig:SP_clusterPartition}).
Subsequently, this clustering structure is incorporated into the GscanStat model with Type IV interaction as part of the divide-and-conquer modeling framework. Specifically, first- and second-order neighborhood models are fitted by partitioning the data into $D=15$ subdomains, defined according to the Autonomous Communities of Spain.
Additionally, a noCluster model is also fitted using the same specifications, serving as a baseline for comparison with the GscanStat model in terms of model selection criteria and relative risk estimates.

\begin{figure}[h!]
   \centering
   \includegraphics[width=0.5\textwidth]{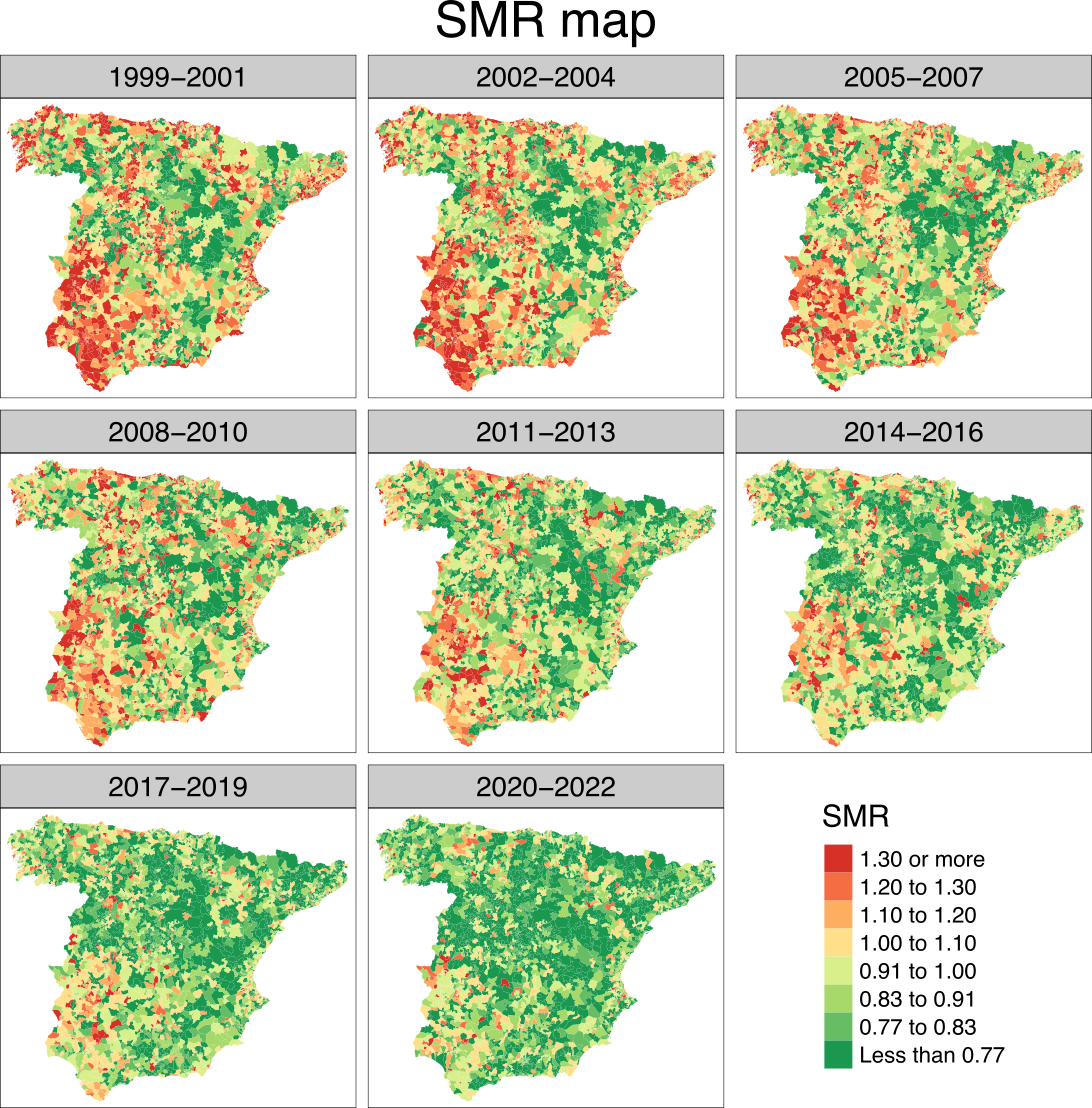}\\[5mm]
   \includegraphics[width=\textwidth]{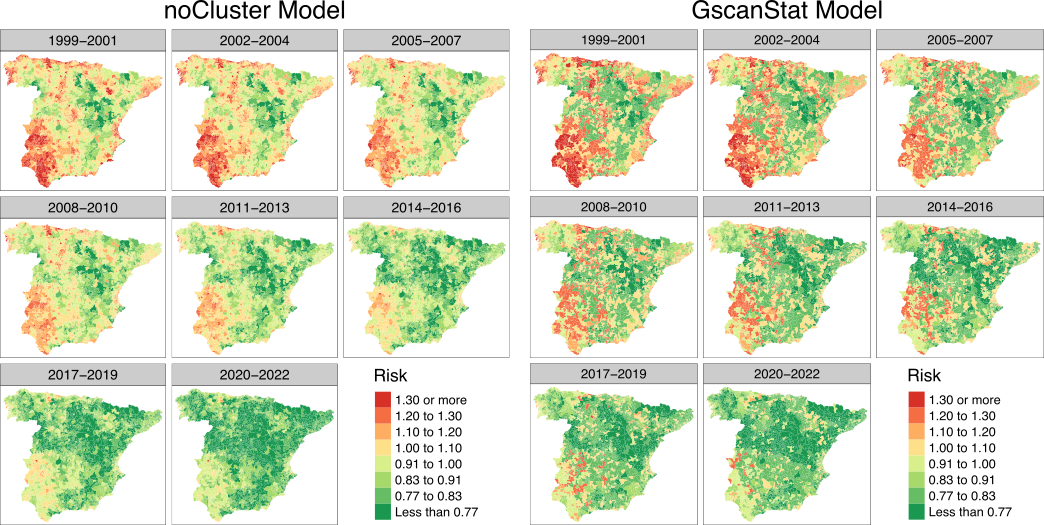}
   \caption{Standardized Mortality Ratios (top) and posterior median estimates of relative risks (bottom) for overall cancer mortality data in Spain.}
   \label{fig:SP_SMR_risks}
\end{figure}

\begin{figure}[h!]
    \centering
    \includegraphics[width=\textwidth]{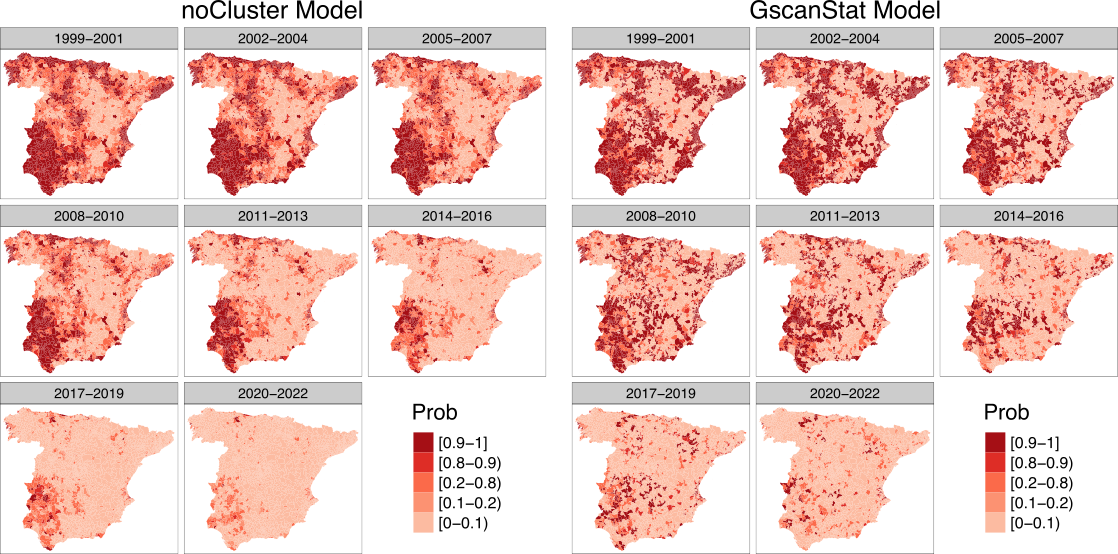}
    \caption{Posterior exceedance probabilities obtained by noCluster and GscanStat models for high-risk areas ($P(r_i^t >1|\bm{O})$).}
    \label{fig:SP_prob_geq_1}
\end{figure}

\autoref{tab:SP_DIC} shows the comparison of fit measures between the model without clusters and the GscanStat-based model. The results shown in the table correspond to first-order neighborhood models within the divide-and-conquer modeling framework, as they outperform models with higher-order neighborhoods. Clearly, incorporating the clustering partition into the model estimation step leads to more accurate estimates of the final relative risks. 
By comparing the raw SMR maps with the risk estimates produced by the noCluster and GscanStat models (see \autoref{fig:SP_SMR_risks}), we observe that the noCluster model tends to oversmooth both high- and low-risk areas, whereas GscanStat preserves finer spatial risk patterns by integrating cluster information.
Despite differences in smoothing, the overall spatio-temporal trends are consistent across both models. During the initial time periods, high-risk areas are primarily concentrated in southwestern Spain (particularly in municipalities within the provinces of Huelva, Sevilla, Cádiz, Badajoz, and Cáceres) as well as along the Cantabrian and North Atlantic coasts. Over successive time periods, these high-risk areas gradually return to baseline or even lower risk levels, reflecting a clear overall decline in mortality rates across Spain during the study period.
This spatio-temporal risk pattern is further corroborated by \autoref{fig:SP_prob_geq_1}, which presents the posterior exceedance probabilities for elevated risk levels.


\begin{table}[!h]
    \centering
    \begin{tabular}{lccccc}
        \hline
        Model & Deviance & $p_D$ & DIC & WAIC\\
        \hline
        noCluster  & 127196.9 & 2895.8 & 130092.6 & 130232.7\\
        GscanStat  & \textbf{120966.0} & \textbf{1923.4} & \textbf{122889.4} & \textbf{122729.7}\\
        \hline
    \end{tabular}
    \caption{Model comparison based on the posterior mean deviance, effective number of parameters ($p_D$), deviance information criterion (DIC), and Watanabe-Akaike information criterion (WAIC).}
    \label{tab:SP_DIC}
\end{table}

\pagebreak

\section{Conclusions}
\label{sec:conclusions}

In this work, we propose a new method that integrates spatio-temporal cluster detection with Bayesian hierarchical disease risk estimation. Our proposal, named \textit{GscanStat}, is a scan-statistic-based method that efficiently identifies arbitrarily shaped high and low-risk clusters using a greedy search strategy. The detected clusters are then incorporated into a Bayesian spatio-temporal model to achieve accurate risk estimation that accommodates local discontinuities.

The extensive simulation study demonstrated that GscanStat outperforms the widely used SaTScan method in cluster detection accuracy. Unlike SaTScan, which relies on a cylindrical scan window and struggles with arbitrarily shaped clusters, our method achieves significantly higher recall and precision, accurately identifying both high- and low-risk clusters while substantially reducing false positives.
Moreover, integrating cluster information into the risk estimation model improves model fit and produces more accurate risk estimates, as evidenced by lower DIC, WAIC, and logarithmic score values compared to both the SaTScan-based model and a standard BYM2 model without clustering information.

GscanStat introduces significant improvements over earlier two-stage modeling approaches, such as \cite{anderson2014,adin2019}, which rely on fitting multiple models to candidate clusters generated by an agglomerative hierarchical algorithm. These agglomerative algorithms are unworkable when applied to a large number of small areas.
In contrast, GscanStat avoids repeated model fitting and efficiently detects clusters of arbitrary shape. A more recent contribution by \cite{santafe2021} proposes a density-based clustering algorithm that facilitates the analysis over large spatial domains. However, this method neither allows for the detection of clusters in spatio-temporal data nor incorporates the underlying data distribution during cluster detection, limitations that GscanStat directly overcomes.

The advantages of GscanStat extend beyond methodological improvements, demonstrating practical utility through its application to cancer mortality data across Spanish municipalities. By capturing meaningful spatio-temporal risk patterns and mitigating the oversmoothing inherent to Bayesian hierarchical models with CAR priors \cite{retegui2025prior}, GscanStat preserves important local variations, which are essential for accurate risk estimation.


In summary, the proposed methodology provides a powerful epidemiological tool for detecting spatio-temporal clusters and estimating disease risk. To facilitate reproducibility and broader application, the \texttt{R} code for implementing the GscanStat clustering algorithm, fitting the partition models using the \texttt{bigDM} package, and reproducing results from the case study is publicly available at \url{https://github.com/spatialstatisticsupna/GscanStat}.

\section*{Acknowledgments}
This work has been supported by project PID2020-113125RB-I00/MCIN/AEI/10.13039/501100011033 (Spanish Ministry of Science and Innovation).

\clearpage
\bibliographystyle{apalike}
\bibliography{bibliography}


\counterwithin*{figure}{section}
\counterwithin*{table}{section}

\renewcommand{\thefigure}{S\arabic{figure}}
\renewcommand{\thetable}{S\arabic{table}}
\setcounter{figure}{0}
\setcounter{table}{0}

\clearpage
\section*{Supplementary material}
\addcontentsline{toc}{section}{Supplementary material}

This supplementary material includes additional tables that complement the simulation study.

\bigskip

\begin{table}[!h]
 \centering
 \begin{tabular}{|c|c|c|c|c|c|c|}
\multicolumn{7}{c}{\bf Average values of metrics across high-risk clusters} \\[1ex]
\hline
\textbf{Scenario} & \textbf{Model} & \textbf{MAB} & \textbf{MRMSE} & \textbf{Length} & \textbf{Cover95} & $\bm{IS_{0.05}}$\\ \hline
\multirow{3}{*}{A-1H}
 & noClusterModel & 0.163 & 1.634 & 0.387 & 0.639 & 1.685\\
 & SaTScan & 0.201 & 2.011 & 0.355 & 0.521 & 3.305\\
 & GscanStat & \textbf{0.031} & \textbf{0.310} & 0.131 & 0.970 & \textbf{0.130}\\
\hline

\multirow{3}{*}{A-1H1L}
 & noClusterModel & 0.153 & 1.539 & 0.406 & 0.697 & 1.215\\
 & SaTScan & 0.179 & 1.803 & 0.382 & 0.595 & 2.389\\
 & GscanStat & \textbf{0.046} & \textbf{0.465} & 0.126 & 0.952 & \textbf{0.127}\\
\hline

\multirow{3}{*}{C-1H}
 & noClusterModel & 0.168 & 1.690 & 0.389 & 0.632 & 1.744\\
 & SaTScan & 0.205 & 2.059 & 0.356 & 0.507 & 3.393\\
 & GscanStat & \textbf{0.048} & \textbf{0.480} & 0.164 & 0.940 & \textbf{0.165}\\
\hline

\multirow{3}{*}{C-1H1L}
 & noClusterModel & 0.155 & 1.554 & 0.408 & 0.692 & 1.229\\
 & SaTScan & 0.186 & 1.859 & 0.384 & 0.584 & 2.418\\
 & GscanStat & \textbf{0.043} & \textbf{0.430} & 0.171 & 0.945 & \textbf{0.172}\\
\hline
\multicolumn{7}{c}{\rule{0pt}{3.5ex} \bf Average values of metrics across low-risk clusters} \\[1ex]
\hline
\textbf{Scenario} & \textbf{Model} & \textbf{MAB} & \textbf{MRMSE} & \textbf{Length} & \textbf{Cover95} & $\bm{IS_{0.05}}$\\ \hline
\multirow{3}{*}{A-1H1L}
 & noClusterModel & 0.236 & 2.367 & 0.814 & 0.737 & 1.938\\
 & SaTScan & 0.271 & 2.723 & 0.738 & 0.654 & 2.357\\
 & GscanStat & \textbf{0.144} & \textbf{1.449} & 0.249 & 0.828 & \textbf{1.910}\\
\hline

\multirow{3}{*}{A-1L}
 & noClusterModel & 0.342 & 3.422 & 0.470 & 0.308 & 5.773\\
 & SaTScan & 0.352 & 3.525 & 0.453 & 0.253 & 6.079\\
 & GscanStat & \textbf{0.063} & \textbf{0.635} & 0.158 & 0.928 & \textbf{1.570}\\
\hline

\multirow{3}{*}{C-1H1L}
 & noClusterModel & 0.231 & 2.310 & 0.796 & 0.727 & 1.939\\
 & SaTScan & 0.267 & 2.667 & 0.721 & 0.647 & 2.372\\
 & GscanStat & \textbf{0.163} & \textbf{1.629} & 0.309 & 0.782 & \textbf{1.742}\\
\hline

\multirow{3}{*}{C-1L}
 & noClusterModel & 0.342 & 3.421 & 0.471 & 0.311 & 5.750\\
 & SaTScan & 0.350 & 3.498 & 0.441 & 0.262 & 6.121\\
 & GscanStat & \textbf{0.089} & \textbf{0.887} & 0.193 & 0.865 & \textbf{1.561}\\
\hline
\end{tabular}
\caption{\label{tab:riskEvaluation_HLClust} Log-risk estimation performance in high/low-risk clusters across 100 simulations.}
\end{table}

\clearpage

\begin{table}[!h]
 \centering
 \begin{tabular}{|c|c|c|c|c|c|c|}

\hline
\textbf{Scenario} & \textbf{Model} & \textbf{MARB} & \textbf{MRRMSE} & \textbf{Length} & \textbf{Cover95} & $\bm{IS_{0.05}}$\\ \hline

\multirow{3}{*}{A-1H}
 & noClusterModel & 0.041 & 0.406 & 0.795 & 0.983 & 0.837\\
 & SaTScan & 0.032 & 0.325 & 0.695 & 0.984 & 0.725\\
 & GscanStat & \textbf{0.018} & \textbf{0.180} & 0.187 & 0.978 & \textbf{0.182}\\
\hline

\multirow{3}{*}{A-1H1L}
 & noClusterModel & 0.045 & 0.447 & 0.868 & 0.984 & 0.899\\
 & SaTScan & 0.036 & 0.365 & 0.787 & 0.983 & 0.808\\
 & GscanStat & \textbf{0.020} & \textbf{0.206} & 0.179 & 0.966 & \textbf{0.177}\\
\hline

\multirow{3}{*}{A-1L}
 & noClusterModel & 0.041 & 0.413 & 0.709 & 0.986 & 0.733\\
 & SaTScan & 0.044 & 0.441 & 0.675 & 0.985 & 0.699\\
 & GscanStat & \textbf{0.021} & \textbf{0.209} & 0.173 & 0.977 & \textbf{0.169}\\
\hline

\multirow{3}{*}{C-1H}
 & noClusterModel & 0.046 & 0.461 & 0.794 & 0.982 & 0.835\\
 & SaTScan & 0.041 & 0.408 & 0.694 & \textbf{0.982} & 0.724\\
 & GscanStat & \textbf{0.035} & \textbf{0.348} & 0.249 & 0.927 & \textbf{0.251}\\
\hline

\multirow{3}{*}{C-1H1L}
 & noClusterModel & 0.051 & 0.512 & 0.867 & 0.984 & 0.896\\
 & SaTScan & 0.046 & 0.459 & 0.784 & 0.983 & 0.805\\
 & GscanStat & \textbf{0.039} & \textbf{0.386} & 0.263 & 0.937 & \textbf{0.261}\\
\hline

\multirow{3}{*}{C-1L}
 & noClusterModel & 0.048 & 0.478 & 0.707 & 0.984 & 0.736\\
 & SaTScan & 0.050 & 0.492 & 0.649 & 0.976 & 0.697\\
 & GscanStat & \textbf{0.040} & \textbf{0.390} & 0.228 & 0.923 & \textbf{0.232}\\
\hline

\end{tabular}
\caption{\label{tab:riskEvaluation_noCluster} Log-risk estimation performance for areas not included in any cluster across 100 simulations.}
\end{table}

\end{document}